\documentclass{JHEP3}

\newcommand{\GG}{{\cal{G}}}

\newcommand{\DD}{\mathcal{D}}   
\newcommand{\WA}{\mathcal{A}}

\def\half{\frac{1}{2}}

\def\alp{\leavevmode\ifmmode {\alpha^\prime} \else ${\alpha^\prime}$ \fi}
\def\GN{G_{N}}

\newcommand{\ud}{\mathrm{d}}

\newcommand{\bea}{\begin{eqnarray}}
\newcommand{\beal}[1]{\begin{eqnarray}\label{#1}}
\newcommand{\eea}{\end{eqnarray}} 
\newcommand{\be}{\begin{equation}} 
\newcommand{\bel}[1]{\begin{equation}\label{#1}}
\newcommand{\ee}{\end{equation}} 
\newcommand{\rf}[1]{(\ref{#1})}
\newcommand{\nn}{\nonumber}
\newcommand{\bit}{\begin{itemize}}
\newcommand{\eit}{\end{itemize}}
\newcommand{\ben}{\begin{enumerate}}
\newcommand{\een}{\end{enumerate}}

\newcommand{\tl}{\theta_\ell}
\newcommand{\tn}{\theta_n}

\def\half{\frac{1}{2}}

\def\alp{\leavevmode\ifmmode {\alpha^\prime} \else ${\alpha^\prime}$ \fi}

\preprint{}

\title{Entropy currents from holography \\ in hydrodynamics with charge}

\author{Grzegorz Plewa\footnote{Email: g.plewa@ipj.gov.pl}\\
National Center for Nuclear Research, ul. Ho\.za 69, 00-681 Warsaw, Poland
}

\author{Micha\l\ Spali\'nski\footnote{Email: mspal@fuw.edu.pl} \\
National Center for Nuclear Research, ul. Ho\.za 69, 00-681 Warsaw, Poland  \\
Physics Department, University of Bia{\l}ystok, ul. Lipowa 41, 15-424 Bia{\l}ystok, Poland 
}

\abstract{ 
The holographic interpretation of the hydrodynamic entropy current is
developed for the case of hydrodynamics with a conserved charge. This is
carried out within a framework developed in earlier work
\cite{Booth:2010kr,Booth:2011qy}, 
which showed how to associate entropy currents with horizons in the dual
geometry. The entropy current defined by the event horizon in the dual bulk
geometry is calculated. It is also shown that to second order in the gradient
expansion the dual geometry possesses a unique Weyl-invariant apparent horizon
which also defines an admissable entropy current. At first order both 
currents coincide with the result obtained on the basis of a purely
hydrodynamic analysis \cite{Son:2009tf}. 
}

\keywords{Gauge/gravity duality, Black Holes, Yang-Mills plasma.}

\begin{document}

\section{Introduction}

For static black holes in Einstein gravity there is a well established notion
of entropy expressed in terms of the event horizon area by the Beckenstein
formula \cite{Bekenstein:1973ur}. One would expect that this notion should, in
some form, apply also to black holes which are in an appropriate sense close
to being static \cite{Booth:2003ji}.  However, the event horizon boundary of a
black hole is a global, teleological concept -- its location can only be
determined once the complete spacetime is known, since one must examine the
ultimate fate of all signals originating from a point before ruling whether or
not that point is part of the black hole \cite{Booth:2005qc}. In this way the
evolution of an event horizon can appear to be acausal. The area of the event
horizon is similarly teleological. For example the area increase of an event
horizon is not directly driven by infalling matter or energy; the actual
effect of an influx through the event horizon is to decrease its rate of
expansion \cite{Booth:2005qc}. This becomes a puzzle as soon as one wishes to
interpret the area of the event horizon as a measure of black hole entropy,
since the apparently acausal expansion of event horizons would then seem to
imply a similarly acausal evolution of entropy. An acute expression of this
comes arises in the context of the AdS/CFT correspondence, which maps the
entropy of an AdS black hole to the entropy of a fluid in the dual quantum
field theory on the conformal boundary.

It is therefore an open question whether a naive reading of the Beckenstein
formula accounts for black hole entropy for dynamical black holes. The
simplest option is to assume that the entropy is still proportional to the
area, but perhaps not to the area of the event horizon itself, but of a
hypersurface which asymptotes to it at late times
\cite{Figueras:2009iu,Booth:2009ct,Booth:2010kr,Booth:2011qy}. One such
example is the apparent horizon, understood as the boundary of a region
containing trapped surfaces. For static black holes the apparent and event
horizons coincide, but the location of an apparent horizon on a given slicing
of spacetime can be determined locally in time. Since the apparent horizon
evolves in a causal way, the associated entropy also has this property. Such a
notion, while not free of conceptual problems, has attracted a lot of
attention \cite{Booth:2005qc}.

In the context of the AdS/CFT correspondence the long wavelength distorsions
of black hole (or black brane) horizons map to hydrodynamic states of the dual
quantum field theory plasma on the AdS conformal boundary
\cite{Bhattacharyya:2008jc}. The notion of near-equilibrium entropy then maps
to the hydrodynamic generalization of entropy -- the entropy current
\cite{Bhattacharyya:2008xc,Loganayagam:2008is}. This notion was introduced in
the framework of relativistic hydrodynamics \cite{LL}. In the perfect fluid
approximation the divergence of the entropy current vanishes on all solutions, but is
non-vanishing already at first order in the gradient expansion.

Within the hydrodynamic framework the entropy current is
constructed phenomenologically in the gradient expansion by requiring that in
equilibrium it reproduces thermodynamic entropy and that its divergence
evaluated on any solution of the equations of hydrodynamics is non-negative. A
detailed analysis of the consequences of this generalized second law of
thermodynamics on the form of the entropy current in
\cite{Romatschke:2009im,Bhattacharyya:2012nq} has showed that already at
second order in gradients there is an ambiguity inherent in such a definition.

It is very natural to suspect that the bulk counterpart of this ambiguity is
related to the choice of horizon assumed to be the carrier. In the case of
hydrodynamics with no conserved charges beyond the energy momentum tensor it
was shown in \cite{Booth:2010kr,Booth:2011qy} that this is indeed the
case. This was done by providing an explicit formula for the entropy current
associated with a given hypersurface satisfying the area theorem\footnote{A
  formula of this type is also discussed in \cite{Compere:2012mt}.} and
asymptoting to the event horizon at late times. This formula reproduces the
result of \cite{Bhattacharyya:2008xc} when evaluated on the event horizon.

From the perspective of the phenomenological definition of the hydrodynamic
entropy current none of these hypersurfaces and none of the available
bulk-boundary maps is favored over any other. However, causality of the
boundary field theory seems to favor the entropy current dual to the apparent
horizon -- provided that it is free of the ambiguities related to foliation
dependence and that the bulk-boundary map in use is causal. These issues were 
discussed at length in \cite{Booth:2010kr}. The present article applies the
same methods to a more complicated, but practically important case of
hydrodynamics with a conserved current. The dual holographic description of
this system was established in \cite{Erdmenger:2008rm,Banerjee:2008th} (see also
\cite{Hur:2008tq,Kalaydzhyan:2010iv}). In
this paper the solution appearing in \cite{Plewa:2012vt} is used, which is somewhat
more general in that it allows for a weakly curved boundary geometry and is
presented in a gauge convenient for the task at hand.

The organization of the paper is the following. Section \ref{SECTintro}
briefly reviews the geometry dual to conformal fluid dynamics with a conserved
current obtained in \cite{Plewa:2012vt}. Section \ref{SECTah} describes the
calculation of the relevant apparent horizon in this geometry up to second
order in gradients. In section \ref{SECTIONec} the formula introduced in
\cite{Booth:2010kr} is used to find and compare entropy currents defined by
the event horizon and by the apparent horizon. A general discussion of the
results and possible future directions of research is provided in Section
\ref{SECTconclusions}. Some technical points are discussed in appendices. 
Appendix \ref{appendixWeyl} provides a very brief
account of Weyl invariance, and appendix \ref{appendixTensors} lists Weyl
invariant tensors needed in the main text. Appendix \ref{appendixHydro}
provides some details of the second order hydrodynamics which follow from the
solution of \cite{Plewa:2012vt}.  Appendix \ref{appendixThetas} contains
explicit results for the expansions needed for the computation of the location
of the apparent horizon. Finally, appendix \ref{appendixFO} relates the results
obtained here to the analysis of reference \cite{Son:2009tf}.

\section{The geometry of fluid-gravity duality with a conserved charge\label{SECTintro}}

To describe hydrodynamics with a conserved current a Maxwell field in the bulk
is required  \cite{Erdmenger:2008rm,Banerjee:2008th,Plewa:2012vt}. 
The action of the five-dimensional Einstein-Maxwell theory under consideration
reads\footnote{Latin indices run over all spacetime dimensions, i.e. from $0$
  to $4$, while Greek indices run over the ``boundary directions'', i.e. from 
  $0$ to $3$.}
\beal{action}
S&=&\frac{1}{16\pi\GN}\int d^5x\sqrt{-g} \left(\frac{12}{L^2}
+ R - F^2 - \frac{4 \kappa}{3} \epsilon^{abcde} A_a F_{bc}F_{de}\right) \, , 
\eea
where $\GN$ is the $5$-dimensional Newton's constant and the cosmological
constant is denoted by $12/L^2$. For this theory to be 
a consistent truncation of type IIB
supergravity\cite{Chamblin:1999tk,Gauntlett:2006ai,Gauntlett:2007ma} the
Chern-Simons 
coupling $\kappa$ has to assume the value $1/2\sqrt{3}$.

The equations of motion
derived from \rf{action} support a 
$5$-parameter family of exact, static black hole solutions
\cite{Erdmenger:2008rm,Banerjee:2008th,Plewa:2012vt} with planar
horizons 
obtained by boosting and dilating the AdS-Reissner-Nordstrom black brane
solution \cite{Chamblin:1999tk}. 
The constant dilation parameter is denoted by $b$, the charge by $q$ and the 
boost parameter $u^{\mu}$ is a $4$-component velocity vector in the ``boundary
directions''\footnote{The $4$-velocity vector is normalized so that  
$u_{\mu} u^{\mu} = -1$ in the sense of the 
metric on the
conformal boundary of the locally asymptotically AdS spacetime \rf{bbb}, whose
components are denoted by 
$h_{\mu \nu}$.}. The solution can be expressed in the form 
\bel{bbb}
\textmd{d} s^2 = r^2 \left(P_{\mu \nu}-2 B  u_{\mu}u_{\nu} \right)
\textmd{d}x^{\mu} \textmd{d} x^{\nu} - 2 u_{\mu} \textmd{d}x^{\mu} \textmd{d} r \, ,
\ee
where\footnote{The notation is chosen so that in the uncharged limit ($q
  \rightarrow 0$) $B$ is equal to $B(b r)$ as defined 
  in \cite{Bhattacharyya:2008mz}.}  
\be
\label{bform}
B = \frac{1}{2} \left( 1-\frac{1}{b^4 r^4} \left( 1 + q^2 b^6 \right)+ \frac{q^2}{r^6}  \right)
\ee
and
\be
P_{\mu \nu} = h_{\mu \nu} + u_{\mu} u_{\nu}
\ee 
is the projector operator onto the space transverse to $u^{\mu}$. The lines of
constant $x^{\mu}$ in \rf{bbb} are ingoing null geodesics, for large $r$
propagating in the direction set by $u^{\mu}$, and the radial coordinate $r$
parametrizes them in an affine way \cite{Bhattacharyya:2008xc}.

The vector potential takes the form:
\bel{ba}
A = \frac{ \sqrt{3} q }{2 r^2} u_{\mu} \textmd{d}x^{\mu} \, .
\ee
The geometry \rf{bbb} may be
regarded as a stack of constant-$r$ $4$-dimensional planes, starting from the
conformal boundary at $r = \infty$ (which is $4$-dimensional Minkowski spacetime), down
to the curvature singularity at $r = 0$. The
latter is shielded by an event horizon at $r = 1/b$. 
The parameter $b$ appearing in \rf{bform} is related
to the Hawking temperature $T$ of the event horizon by 
\be
T=\frac{1}{2 \pi b} (2 - q^2 b^6)\, .
\ee
It will be assumed that the charge $q$ is always below the extremal
limit, so that $ q^2 b^6 < 2$.
 
If $b$, $q$, $u^{\mu}$ and $h_{\mu \nu}$ are allowed to vary
slowly compared to 
the scale set by $b$, the metric \rf{bbb} should be an
approximate solution of nonlinear Einstein's equations, with corrections
organized in an 
expansion in the number of gradients in the ``boundary directions'' 
parametrized by $x^\mu$. 
This gradient-corrected metric can be written in the form
\bel{wimetric}
\textmd{d} s^2 = \left( {\cal{G}}_{\mu \nu} - 2 u_{\mu} {\cal{V}}_{\nu} \right)
\textmd{d}x^{\mu} \textmd{d} x^{\nu} - 2 u_{\mu} \textmd{d}x^{\mu}
\left(\textmd{d} r  + r \WA_{\nu} \ud  x^{\nu}\right)\, ,
\ee
with the condition $u^{\mu} \GG_{\mu \nu} = 0$  completely fixing the gauge
freedom \cite{Bhattacharyya:2008mz}. 
The field $\WA$ appearing in \rf{wimetric} is the ``Weyl
connection'' defined in appendix \ref{appendixWeyl}. 
It is clear that lines of constant
$x$ are geodesics (affinely parametrized by $r$), as
in the case of the static metric \rf{bbb}. 

The form of the metric \rf{wimetric} is strongly restricted by the
Weyl-invariance\footnote{A brief account of Weyl invariance appears in
  appendix \ref{appendixWeyl}.} of
the bulk theory \cite{Bhattacharyya:2008mz}: the
functions 
$\cal{V_{\mu}}$  are of unit Weyl weight and ${\cal{G}}_{\mu \nu}$ are Weyl
invariant. They can be expressed as linear combinations of independent Lorentz vectors and
tensors built out of $b$, $q$, $u^\mu$, $h_{\mu\nu}$ and their derivatives,
order by order in the 
gradient expansion. 
At first order in gradients one has a Weyl-invariant vector
\be
l_{\mu} = \epsilon_{\mu\nu\lambda\rho} u^\nu \DD^\lambda u^\rho \, ,
\ee
pseudovector
\be
{V_{0}}_\mu = q^{-1} P_{\mu}^{\nu} \DD_{\nu}{q}  \, , 
\ee
and a second order symmetric tensor\footnote{Symmetrization is defined
  as $A_{(\mu \, \nu)}:= A_{\mu \nu}+ A_{\nu \mu}$.}
\be 
\sigma_{\mu \nu} = \half \DD_{(\mu} u_{\nu)}
\ee
of Weyl weight $-1$. The independent objects appearing at second order in gradients are
listed in appendix \ref{appendixTensors}. 

As shown in \cite{Plewa:2012vt}, the resulting solution, up to second order, takes the
form  
\bea
 \label{solMetric}
 \nonumber
 {\cal{V}}_{\mu} &=& r^2 B u_{\mu} + r {F_1} l_{\mu}+b r^2 {F_0}{V_0}_{\mu}  + 
  r^2 \sum_{i=1}^{6} K_{i} S_{i}u_{\mu} + r \sum_{i=1}^{5} W_{i} {V_{i}}_{\mu}
  \, ,
 \\[0ex]
 {\cal{G}}_{\mu \nu} &=& r^2 P_{\mu \nu} + 2 b r^2 F_2 \sigma_{\mu \nu} + 
 r^2 \sum_{i=1}^{6} L_{i} S_i P_{\mu \nu} + \sum_{i=1}^{11}  H_{i} {T_i}_{\mu
   \nu} \, .
 \eea
The $31$ coefficient functions  $F_0,F_1,F_2$, $K_i, L_i$ ($i=1,...,6$), $W_i$
($i=1,...,5$), and ${H_i}$ ($i=1,...,11$) all depend on the Weyl invariant
variables $b^3 q$ and $b r$. 

The Maxwell gauge field $A$ is a vector field
of Weyl weight zero.  It will be taken in the gauge $A_r=0$ and its form is
\cite{Plewa:2012vt}  
\bel{solGauge}
A = \left(   \frac{   \sqrt{3} q u_{\mu}   }{2 r^2} + {Y_0} l_{\mu}+ 
{\tilde{Y}_0} {V_0}_{\mu} +
 r \sum_{i=1}^{6} N_{i} S_i u_{\mu}   +  \sum_{i=1}^{5} Y_{i}{V_i}{_\mu}  
 \right) \textmd{d}x^{\mu} \, .
 \ee
As before, the coefficient functions  $\tilde{Y}_0$, $Y_0,\dots Y_5$, $N_1,
\dots, N_6 $ depend on the Weyl invariant variables $b r$ and $b^3 q$.

All the coefficient functions appearing in \rf{solMetric}, \rf{solGauge} have
been determined by solving the equations of motion\footnote{This, as well as
  most other computations in this paper, was done with the help of
  Mathematica.} and are given explicitly in \cite{Plewa:2012vt}.

The metric given above is a solution of the Einstein-Maxwell equations with
negative cosmological constant up to second order in gradients, provided that
$b$, $q$ and $u^{\mu}$ satisfy the equations of hydrodynamics, i.e. the
equations of covariant conservation of the energy-momentum tensor and charge
current obtained from \rf{solMetric} and \rf{solGauge} by holographic
renormalization \cite{deHaro:2000xn}. Explicit formulae \cite{Plewa:2012vt}
can be found in appendix \rf{appendixHydro}.

\section{The Weyl-invariant apparent horizon \label{SECTah}}   

This section is devoted to locating the Weyl-invariant apparent horizon for
the spacetimes defined by the metric \rf{wimetric} following the approach of
\cite{Booth:2010kr,Booth:2011qy}. The time-evolved apparent horizon, denoted by
$\Delta$, is presented as the level set of a scalar function $S(r,x)$. This
function is required to be Weyl-invariant and can be written in the gradient
expansion in terms of all the independent scalars available up to some
order. Since the outer event horizon at order zero is at $r=1/b$ and at that
order $\Delta$ should coincide with it, one has
\bel{formofs}
S(r,x) = b(x) r - g(x) \, ,
\ee
where $g(x)$ is a Weyl-invariant function expanded in gradients of $b, q$: 
\be
g(x) = 1 + g_1(x) + g_2(x) \ + \dots \, .
\ee
Here $g_k$ denotes a linear combination of all Weyl-invariant scalars at
order $k$ in the gradient expansion. There are no Weyl-invariant scalars at
order 1, and 6 at order 2, so one expects to find 
\bea
\nonumber
g_1(x) &=& 0 \, , \\
g_2(x) &=& \sum_{k=1}^{6} h_k(b^3 q) S_k \, ,
\eea
where the $S_i$ are the $6$ independent Weyl-invariant scalars \rf{scalars}. 
The functions\footnote{In principle these functions also depend on the other
  Weyl-invariant scalar $br$, but these formulae are eventually
  evaluated on hypersurfaces $r=1/b$ up to terms of second order in
  gradients. For this reason one can set $b r=1$  in these functions, since
  they always appear multiplied by second order scalars.} $h_i$ will be  
determined in due course by solving the condition $\tl=0$. Once this is done,
the expression for the  
position of the apparent horizon will take the form
\bel{horpos}
r_H = \frac{1}{b} \left(1 +  \sum_{k=1}^{6} h_k S_k 
\right) \, . 
\ee
The approach developed in \cite{Booth:2010kr,Booth:2011qy} starts by
determining a vector field, denoted by $v$, which is normal to the outer marginally trapped
surfaces which foliate the horizon $\Delta$. The two properties of $v$ which 
are essential to its construction are the fact that $v$ is tangent to 
$\Delta$, and that it must be surface forming, that is, it must satisfy the
Frobenius condition 
\bel{frob}
v \wedge dv = 0 \, . 
\ee
It will be shown in the following that
for the geometry under consideration, up to second order in the gradient
expansion, these conditions together with Weyl invariance determine $v$
uniquely. 

To find the vector field $v$, one first needs the normal covector
to a surface of the form \rf{formofs}. This is given by $m = dS$, 
which up to second order in the gradient expansion 
can be written in terms of the Weyl-covariant 
derivatives (given in \rf{Db}, \rf{Dq}) as 
\be
m =  r \DD_\mu b\, dx^\mu  +  b \left( dr + r \mathcal{A}_\mu dx^\mu\right) \, . 
\ee
The vector $v$ is to be tangent to $\Delta$, so it must satisfy $v\cdot m =
0$. To solve this condition it is convenient to make a special
coordinate choice. As discussed in detail in \cite{Booth:2011qy}, given a
parametrization 
$y^\mu$ of the horizon $\Delta$ (such that $S(r(y), x(y)) \equiv const$), one can choose
the gauge $y=x$ and then the vector field $v$ takes the form 
\bel{genv}
v =  v^\beta 
\left\{
\frac{\partial}{\partial x^\beta} - 
\left(
\frac{\partial S}{\partial r}
\right)^{-1} 
\left(
\frac{\partial S}{\partial x^\beta}
\right)
 \frac{\partial}{\partial r} 
\right\} \, .
\ee
Requiring that the vector $v$ be Weyl invariant implies that (up to second
order) the ``boundary'' components $v^\mu$ take the form
\bel{vsecord}
v^{\mu} = b \left(u^\mu + b \left(a_0 {V_0}^\mu + {a_1}  l^\mu + 
 \sum_{k=1}^5 c_k V_k^\mu \right) + u^{\mu} \sum_{k=1}^6 e_k S_k\right) \, ,
\ee
where $a_k$, $c_k$,  $e_k$ are some functions of the Weyl-invariant
combination $b^3 q$. Using \rf{genv} one finds that the $r$ component of $v$ is
\be
v^r = -b r   A_{\mu } u^{\mu }-r  u^{\mu } \DD_{\mu }b 
- b r \left(b  A_{\mu } +  \DD_{\mu }b \right) \left(a_0 {V_0}^{\mu } + a_1  l^{\mu } \right) \,.
\ee
It is computationally convenient to 
normalize $v$ so that 
\bel{mvnormal} 
m^2 + v^2 = 0  \, ,
\ee 
which implies in particular that coefficients of the longitudinal terms in
\rf{vsecord} 
vanish: $e_k=0$ for $k=0,\dots,6$. 

The remaining coefficient functions appearing in $v$ are also not 
arbitrary. As discussed earlier, to ensure that the vector $v$ defines a
foliation one 
has to impose the Frobenius condition \rf{frob}. 
As was the case in the analysis of \cite{Booth:2011qy}, the vanishing of
$v_{[\mu} \partial_\nu v_{\rho]}$ is 
automatic, but the conditions $v_{[r} \partial_\nu v_{\rho]}=0$ are
nontrivial. In fact these conditions determine the remaining
freedom in $v$. At first order in the gradient expansion one finds
\bea
\nonumber
{a_0} &=& -\frac{3 b^6 q^2 \left(b^6 q^2+2\right)}{4 \left(b^{12} q^4-b^6 q^2-2\right)},
\\[1ex]
{a_1} &=& -\frac{\sqrt{3} b^9 \kappa  q^3 }{b^6 q^2+1} \, ,
\eea
while at second order one obtains
\beal{vcoeffs}
 c_1 &=& \frac{1}{b^6 q^2-2}-{W_1}\left(b^3 q,1\right),
 \nonumber \\
 c_2 &=& \frac{1}{2}-\frac{6 b^{12} q^4 \kappa^2}{(1+b^6 q^2)^2},
 \nonumber \\
 c_3 &=& \frac{2 \sqrt{3} b^9 \kappa  q^3 {F_2}\left(b^3 q,1\right)}{b^6 q^2+1}-{W_3}\left(b^3 q,1\right),   \nonumber\\
   c_4 &=& \frac{3 b^6 q^2}{\left(b^6 q^2-2\right)^2}+\frac{3 b^6 q^2 \left(b^6 q^2+2\right) {F_2}\left(b^3 q,1\right)}{2 \left(b^6 q^2-2\right) \left(b^6 q^2+1\right)}-{W_4}\left(b^3 q,1\right),   \nonumber\\
   c_5 &=& - {W_5}\left(b^3 q,1\right).
\eea
To show that the Frobenius condition \rf{frob} is really satisfied one also has
to use the relations  
\beal{fromhydro}
\partial_\mu \WA_\nu - \partial_\nu \WA_\mu &=& 0 \, , \nn\\
\DD_\mu {V_0}_\nu - \DD_\nu {V_0}_\mu &=& 0 
\eea
(valid up to second order in gradients). These relations 
(discussed further in appendix \ref{appendixHydro})  
follow from the equations of
hydrodynamics. Since the Frobenius
condition was imposed for the full spacetime (rather than just on the
horizon), the vector $v$ actually gives rise to a foliation of the full
spacetime, at least in a neighhborhood of $\Delta$.

This way one finds (as in the uncharged case \cite{Booth:2011qy}) that the
foliation vector $v$ is completely determined once $\Delta$ is fixed. As
discussed in \cite{Booth:2011qy}, it seems plausible that this will also be
the case at higher orders in the gradient expansion.  

Dynamical quasilocal horizons are spacelike and so $m$ should be
timelike and $v$ spacelike. Without loss  
of generality one can assume that $m$ is future oriented and $v$ is outward pointing. 
Then the null normals to the leaves of the foliation of $\Delta$ can be
expressed as 
\beal{evos}
v &=& \ell - C n \nn \\
m &=& \ell + C n \, ,
\eea
where the scalar $C$ is called the evolution parameter
\cite{Booth:2003ji,Booth:2006bn}: 
\bel{cform}
C = \half v^2 \, .
\ee
The sign of the evolution parameter indicates whether $\Delta$ is
spacelike ($C>0$), timelike ($C<0$) or null ($C=0$). The signs of the coefficients in
\rf{evos} have been chosen to 
ensure that both $\ell$ and $n$ are future oriented, and $\ell$ is
outward-pointing while $n$ is inward-pointing. Given the formulae for $m$ and
$v$ it is straightforward to  
calculate $C$, $\ell$ and $n$ explicitly. 

To determine the position of the apparent horizon one needs to calculate the
null expansions 
\bea
\nonumber
\tl &=& \tilde{q}^{ab} \nabla_a \ell_b\,  , \\
\tn &=& \tilde{q}^{ab} \nabla_a n_b \, ,
\eea
where 
\be
\tilde{q}_{ab} = G_{ab} + \ell_a n_b + \ell_b n_a 
\ee
is the metric induced on the foliation slices and $G_{ab}$ are the components
of the metric \rf{wimetric}.  
Using the results of the previous section one finds (up to
second order)\footnote{This computation is fairly lengthy.} 
\beal{thetares}
\tl &=&  3 b B r  + \sum_{k=1}^6 \tl^{(k)} S_k \, , \\
\tn &=&  -\frac{3}{ b r} + \sum_{k=1}^6 \tn^{(k)} S_k\, ,
\eea
where the coefficient functions $\tl^{(k)}$ and  $\tn^{(k)}$ are listed in 
appendix \ref{appendixThetas}. 
Note that the results are manifestly Weyl-invariant. There is
no correction at first order, as required by Weyl invariance.   
With these results in hand, it is straightforward to determine the location of
the apparent horizon by solving $\tl(r_{AH}) = 0$. One finds the result 
\rf{horpos} with 
\beal{AH}
\nonumber
h_1^{AH} &=& \frac{{K_1}\left(b^3 q,1\right)}{b^6 q^2-2},
\\[1ex]
h_2^{AH} &=& \frac{9-2 b^6 q^2}{12 \left(b^6 q^2-2\right)}+\frac{2 b^{12} \kappa ^2 q^4}{5 \left(b^6 q^2+1\right)^2} ,
\nonumber
\\[1ex]
h_3^{AH} &=& \frac{1}{12 \left(b^6 q^2-2\right)},
\nonumber
\\[1ex]
h_4^{AH} &=& \frac{ {K_4}\left(b^3 q,1\right)}{b^6 q^2-2}+\frac{b^6 q^2
  \left(b^{24} q^8-102 b^{18} q^6-244 b^{12} q^4-232 b^6 q^2-64\right)}{32
  \left(b^6 q^2-2\right)^4 \left(b^6 q^2+1\right)^2}, 
\nonumber
\\[1ex]
h_5^{AH} &=& \frac{ {K_5}\left(b^3 q,1\right)}{b^6 q^2-2}-\frac{b^6 q^2
  \left(b^6 q^2+2\right)}{4 \left(b^6 q^2-2\right)^2 \left(b^6 q^2+1\right)}, 
\nonumber
\\[1ex]
h_6^{AH} &=& \frac{ {K_6}\left(b^3 q,1\right)}{b^6 q^2-2}+\frac{\sqrt{3} b^9
  \kappa  q^3 \left(3 b^{12} q^4+14 b^6 q^2+8\right)}{4 \left(b^6
  q^2-2\right)^2 \left(b^6 q^2+1\right)^2} \, .
\eea
In the uncharged case considered in \cite{Booth:2010kr,Booth:2011qy}, only $h_1$ differs
from the result for the event horizon. In the present case, the
event horizon determined in \cite{Plewa:2012vt} differs from the apparent horizon in
the coefficients $h_1$ and $h_4$: 
\bea
h_1^{(EH)} &=& h_1^{(AH)} + \frac{1}{3 (b^6 q^2-2)^2} \\
h_4^{(EH)} &=& h_4^{(AH)} + \frac{b^6 q^2 \left(b^6 q^2+2\right)^2}{2
  \left(b^6 q^2-2\right)^4} \, .
\eea
The expression 
\bel{horsep}
r_{EH} - r_{AH} =  \frac{1}{b} \frac{1}{\left(b^6 q^2-2\right)^2}  \left(\frac{1}{3} S_1 + 
\frac{b^6 q^2}{2}  \frac{\left(b^6 q^2+2\right)^2}{\left(b^6 q^2-2\right)^2} S_4 \right)
 \geq 0
\ee
explicitly shows\footnote{Both $S_1$ and $S_4$ are manifestly positive.} that
the apparent horizon lies within (or coincides with) the 
event horizon in the sense that an ingoing radial null geodesic will cross first the
event horizon and only then the apparent horizon, since $r$ is an affine
parameter on such geodesics. It is also easy to check that (in accordance with
expectations) the apparent
horizon is spacelike or null 
\be
C(r_{AH}) =  \frac{1}{\left(2 - b^6 q^2\right)} \left(
\frac{1}{3}  S_1 +
\frac{b^6 q^2}{2}  \frac{\left(b^6 q^2+2\right)^2}{\left(b^6 q^2-2\right)^2} S_4
\right) \geq 0 \, , 
\ee
as long as the black brane is subextremal (that is, $b^6 q^2 < 2$).

\section{Hydrodynamic entropy currents\label{SECTIONec}}  

In hydrodynamics the entropy current is a phenomenological notion constructed
order-by-order in the gradient expansion. The leading term describes the
flow of thermodynamic entropy in the perfect fluid approximation. In the case
considered in this paper, the most 
general form of the entropy current consistent with conformal symmetry up to 
second order in gradients reads
\bel{ecres}
S^\mu =  \frac{1}{4 \GN} b^{-3} \left(
u^\mu  + b \left( j_0  V_0^{\mu} + j_1  l^{\mu }  + \sum_{k=1}^5 j^\perp_k
V_{k}^{\mu} \right) +  \left(\sum_{k=1}^6 j^{||}_k S_{k}\right) u^\mu 
\right) \, .
\ee
The overall factor of $1/4 \GN$ in \rf{ecres} comes from the holographic
formula for thermodynamic entropy. The coefficients $ j_0$, $j_1$, $j^\perp_k$
and 
$j^{||}_k$ appearing in this expression are functions of the 
Weyl-invariant variable $b^3 q$ and should be such as to ensure that the
divergence of this current is non-negative on all solutions of the equations
of hydrodynamics. For the case without charge this condition was analyzed by
\cite{Romatschke:2009kr}, where it was found that it is not possible to
determine all these coefficients in terms of the transport coefficients without
some additional input.  
Thus, if the notion of local entropy
production in the near-equilibrium regime makes sense, there must be some
further constraints on 
the form of the hydrodynamic entropy current. 

This prompted
the analysis of \cite{Booth:2010kr}, which showed that the ambiguity in the entropy 
current is reflected on the gravity side precisely as an ambiguity in the
choice of horizon used to define the entropy current. In \cite{Booth:2010kr} a formula
was proposed, which associates an entropy current with each hypersurface which
satisfies the Hawking area increase theorem, such as the event or apparent
horizons: 
\bel{entrocur}
S^\mu = \frac{1}{4 \GN} \frac{1}{b} \sqrt{\frac{G}{h}}\ v^\mu \, ,
\ee
where $G$ is the determinant of the bulk metric \rf{wimetric} and $h$ is the
determinant of the weakly 
curved boundary metric. 
This formula, as discussed at
length in \cite{Booth:2010kr}, is to be evaluated on the chosen horizon. The divergence
of this current was shown to be proportional to the quantity $\tl-C\tn$, which
is non-negative on the basis of the area theorem. 

To evaluate the entropy current according to \rf{entrocur} one needs to calculate the
determinant of the metric 
\rf{wimetric}. This determinant is proportional to $h$; up to terms of higher order in the
gradient expansion one finds
\be
\frac{G}{h} = r^6 \left(1 + 
\left(3 {L_1}-2 {F_2}^2\right) {S_1} +
  \left(\frac{1}{b^2 r^2}-\frac{12 b^6 \kappa ^2 q^4}{5 r^6 \left(b^6
    q^2+1\right)^2}\right) {S_2} 
+  3 \left({L_4} {S_4}+{L_5} {S_5}+ {L_6} {S_6}  \right)\right) \, .
\ee
Using this, as well as the vector $v$ determined earlier, one can evaluate \rf{entrocur}
on a hypersurface of the form \rf{horpos}. This  
leads to the result \rf{ecres}. Coefficients of the first order terms are
found to be 
\beal{econe}
j_0 &=& - \frac{3 b^6 q^2   \left(b^6 q^2+2\right) }{4 \left(b^6 q^2-2\right)
  \left(b^6 q^2+1\right)}, \nn\\
j_1 &=&  -  \frac{\sqrt{3} b^{9} \kappa  q^3   }{b^6 q^2+1} \, .
\eea
Coefficients of the transverse second order terms are given by
\beal{jperpcoeffs}
\nonumber
j_1^{\perp} &=&  \frac{1}{b^6 q^2-2} -  {W_1}\left(b^3 q,1\right)\, ,
\\[1ex]
\nonumber
j_2^{\perp} &=& \frac{1}{2}-\frac{6 b^{12} \kappa ^2 q^4}{\left(b^6 q^2+1\right)^2} \, ,
\\[1ex]
\nonumber
j_3^{\perp} &=& \frac{2 \sqrt{3} b^9 \kappa  q^3 {F_2}\left(b^3
  q,1\right)}{b^6 q^2+1} - {W_3}\left(b^3 q,1\right) \, ,
\\[1ex]
\nonumber
j_4^{\perp} &=& \frac{3 b^4 q^2}{\left(b^6 q^2-2\right)^2}+\frac{3 b^4 q^2
  \left(b^6 q^2+2\right) {F_2}\left(b^3 q,1\right)}{2 \left(b^6 q^2-2\right)
  \left(b^6 q^2+1\right)}  - {W_4}\left(b^3 q,1\right) \,  ,
\\[1ex]
j_5^{\perp} &=& - {W_5}\left(b^3 q,1\right) \, .
\eea
These are all fixed independently of the hypersurface on which
the formula \rf{entrocur} is evaluated. The coefficient functions of the
longitudinal terms, 
proportional to the 4-velocity $u^\mu$, do however depend on the choice of horizon 
through the functions $h_k$: 
\beal{jparcoeffs}
\nonumber
j_1^{\parallel} &=& -{F_2}\left(b^3 q,1\right)^2+\frac{3}{2} {L_1}\left(b^3
q,1\right)+3  {h_1}\, ,
\\[1ex]
\nonumber
j_2^{\parallel} &=& \frac{1}{2} - \frac{  6 b^{12} q^4 \kappa^2  }{5(1+b^6
  q^2)^2} +3  {h_2}\, ,
\\[1ex]
\nonumber
j_3^{\parallel} &=& 3  {h_3} \, ,
\\[1ex]
\nonumber
j_4^{\parallel} &=& \frac{3}{2} {L_4}\left(b^3 q,1\right)+3  {h_4} \, ,
\\[1ex]
\nonumber
j_5^{\parallel} &=& \frac{3}{2}  {L_5}\left(b^3 q,1\right)+3  {h_5} \, ,
\\[1ex]
j_6^{\parallel} &=& \frac{3}{2} {L_6}\left(b^3 q,1\right)+3  {h_6} \, . 
\eea
Evaluating \rf{entrocur} on the event horizon found in \cite{Plewa:2012vt} leads to the
result 
\beal{eceh}
\nonumber
j_{1}^{\parallel} &=& -{F_2}\left(b^3 q,1\right)^2 + \frac{3}{2}
{L_1}\left(b^3 q,1\right)+\frac{3 {K_1}\left(b^3 q,1\right)}{b^6
  q^2-2}+\frac{1}{\left(b^6 q^2-2\right)^2}, 
\\[1ex]
\nonumber
j_{2}^{\parallel} &=& \frac{5}{4 b^6 q^2-8} \, ,
\\[1ex]
\nonumber
j_{3}^{\parallel} &=& \frac{1}{4 b^6 q^2-8} \, ,
\\[1ex]
\nonumber
j_{4}^{\parallel} &=& \frac{3}{2} {L_4}\left(b^3 q,1\right)+\frac{3 {K_4}\left(b^3 q,1\right)}{b^6 q^2-2}+\frac{3 b^{12} q^4 \left(17 b^{12} q^4+28 b^6 q^2+20\right)}{32 \left(b^6 q^2-2\right)^3 \left(b^6 q^2+1\right)^2} \, ,
\\[1ex]
\nonumber
j_{5}^{\parallel} &=& -\frac{3 b^6 q^2 \left(b^6 q^2+2\right)}{4 \left(b^6 q^2-2\right)^2 \left(b^6 q^2+1\right)} +
\frac{3}{2} {L_5}\left(b^3 q,1\right)+\frac{3 {K_5}\left(b^3 q,1\right)}{b^6 q^2-2} \, ,
\\[1ex]
j_{6}^{\parallel} &=& \frac{3 \sqrt{3} b^9 \kappa  q^3 \left(3 b^{12} q^4+14 b^6 q^2+8\right)}{4 \left(b^6 q^2-2\right)^2 \left(b^6 q^2+1\right)^2} +
\frac{3}{2} {L_6}\left(b^3 q,1\right)+\frac{3 {K_6}\left(b^3 q,1\right)}{b^6 q^2-2} \, .
\eea
For the apparent horizon one finds the same results as above apart from the
coefficient 
functions $j_{1}^{\parallel}$ and $j_{4}^{\parallel}$:
\bea
j_{1}^{\parallel (EH)} - j_{1}^{\parallel (AH)} &=& \frac{1}{\left(b^6
  q^2-2\right)^2} \, , \nn\\
j_{4}^{\parallel (EH)} - j_{4}^{\parallel (AH)} &=& 3 \frac{b^6 q^2}{2}
\frac{\left(b^6 q^2+2\right)^2}{\left(b^6 q^2-2\right)^4} \, .
\eea
In the limit of vanishing charge one can check that the results of
\cite{Booth:2010kr,Booth:2011qy} 
are reproduced\footnote{Note however that in the published version of
  \cite{Booth:2011qy} there is a sign error in eq. (81).}. 

The fact that the
two entropy currents considered here differ in the 
coefficients of both $S_1$ and $S_4$ suggests that in this case there is a two
parameter family of entropy currents satisfying all the hydrodynamic
criteria. In the case without charge there was only a one parameter
family\footnote{Only parameters which appear in the divergence of the entropy
  current appear to be significant, and only such are considered in this
  counting.} -- 
this was shown by hydrodynamic arguments by Romatschke
\cite{Romatschke:2009kr}, but the generalization of this reasoning to
hydrodynamics with charge is not known at present. 

The second order contribution to the entropy current cannot be expressed
uniquely in terms of the transport coefficients. At first order however there
is no ambiguity, and indeed the result obtained above is unique at that
order. One can easily see that this result is consistent with the purely
hydrodynamic analysis of Son and Sur\'owka \cite{Son:2009tf} (see also
\cite{Sadofyev:2010pr}), who expressed the first order entropy current in
terms of the transport coefficients. This is discussed further in appendix
\ref{appendixFO}.

According to the analysis of \cite{Booth:2010kr}, the divergence of the
entropy current is proportional to the quantity $\tl - C \tn$ evaluated on the
relevant horizon. Calculating this expression on either one of the horizons up
to second order in gradients, one obtains
\be
\tl - C \tn\big |_{r=r_{AH}}  =    \frac{3}{\left(2 - b^6 q^2\right)} \left(
\frac{1}{3} S_1 + \frac{b^6 q^2}{2}  \frac{\left(b^6 q^2+2\right)^2}{\left(b^6
  q^2-2\right)^2} S_4 \right)
 \, ,
\ee
which matches the hydrodynamic result. Thus, 
as expected on the 
basis of the area theorem, 
the divergence of both entropy
currents considered here is non-negative, assuming that the charge remains
sub-extremal (i.e. $ b^6 q^2<2$).

It is interesting to observe that the second order divergence of the entropy
current computed above is proportional to the separation of the event and
apparent horizons \rf{horsep}. This is not an accident, since the event
horizon coincides with the apparent horizon in equilibrium, so the divergence
of the entropy current has to vanish when the separation of the two horizons
goes to zero. The second order divergence of the entropy current must
therefore be proportional to the separation computed to second order. At
higher orders it must still be true that the divergence is proportional to the
separation, but it is not clear whether the coefficient would be a zeroth
order quantity.

The divergence of the full second order entropy current is of third order in
gradients, so calculating it by evaluating the quantity  $\tl - C \tn$
(by a bulk computation) would
require determining the geometry to third order in gradients, which has so far
not been done.

\section{Summary\label{SECTconclusions}}

The ambiguity in the definition of the entropy current in relativistic second
order hydrodynamics was recently connected with the issue of dynamical black
hole boundaries \cite{Booth:2010kr}. This
was based on the holographic representation of strongly coupled supersymmetric
Yang-Mills plasma in the framework of fluid-gravity duality
\cite{Bhattacharyya:2008jc,Bhattacharyya:2008xc}. Reference
\cite{Booth:2010kr} proposed an 
explicit formula which associates an entropy current with a horizon in the
bulk geometry. As long as the horizon satisfies the area theorem, the
divergence of the corresponding hydrodynamic entropy current is non-negative.  

In references \cite{Booth:2010kr,Booth:2011qy} entropy currents defined by the
event horizon and the Weyl-invariant apparent horizon were considered in the
case of conformal hydrodynamics with no conserved charges beyond the
energy-momentum tensor.  This line of research was continued here by studying 
the holographic representation of entropy currents in the case of
hydrodynamics with a conserved charge
\cite{Erdmenger:2008rm,Banerjee:2008th,Plewa:2012vt}.  As in
\cite{Booth:2010kr,Booth:2011qy}, currents associated with two horizons were
considered: the event horizon (whose location in the relevant geometry
was established in \cite{Plewa:2012vt}) and the Weyl-invariant apparent
horizon, which was found in this paper using the novel approach to locating
apparent horizons developed in \cite{Booth:2011qy} . This method is
particularly suited to situations, where an apparent horizon which respects a
given symmetry is sought. The work reported here provides another (much more
complex) example where this approach can successfully be applied.

The key idea behind \cite{Booth:2011qy} is that the apparent horizons of
interest in the context of fluid-gravity duality are only those which are
covariant in the sense of the dual hydrodynamic description. In the case
considered here this means they are specified 
covariantly (in the boundary sense) in terms of $b$, $q$, $u^{\mu}$
and their gradients. An important further constraint in the 
present context is Weyl
invariance.

One of the main results of this paper is that up to second order in gradients
a unique Weyl-invariant apparent horizon exists which is covariant in this 
hydrodynamic sense\footnote{The arguments given in \cite{Booth:2011qy}
  suggesting that the uniqueness of the Weyl-invariant apparent horizon
  persists at higher orders apply also here.}.  As in the case without a
conserved charge, this horizon is isolated at leading and first subleading
orders of the gradient expansion and becomes spatial once second order
gradient contributions are included. As in \cite{Booth:2010kr}, this apparent
horizon gives rise to a notion of hydrodynamic entropy current which satisfies
all the hydrodynamic constraints\footnote{The issue of the bulk-to-boundary
  map, introduced in \cite{Bhattacharyya:2008xc} was further discussed in
  \cite{Booth:2010kr,Booth:2011qy}. The present work brings nothing new in
  this regard.}.

At first order in gradients the entropy current in hydrodynamics with charge
was considered by Son and Sur\'owka \cite{Son:2009tf}. They carefully analyzed
hydrodynamic constraints, allowing for the possibility of a $U(1)$ anomaly,
and succeeded in expressing the entropy current at first order in terms of the
transport coefficients. As shown in section \ref{SECTIONec}, the result of
this purely hydrodynamic analysis is reproduced by the holographic entropy
current formula proposed in reference \cite{Booth:2010kr}.

At second order, in the case of conformal hydrodynamics without charge, a
purely hydrodynamic analysis of the allowed form of entropy currents was
performed in \cite{Romatschke:2009kr} (see also
\cite{Bhattacharyya:2012nq}). The current associated with the apparent horizon
was found to be consistent with that analysis \cite{Booth:2010kr}.  In the
case with charge such an analysis has not yet been carried out; it would
furnish an nontrivial check on the results obtained in this paper, which imply
that in hydrodynamics with a charge current there should be at least a two
parameter family of entropy currents at second order in gradients.

It would also be very interesting to study apparent horizons in
the cases of non-conformal \cite{Kanitscheider:2009as}, and superfluid
\cite{Bhattacharya:2011eea,Herzog:2011ec} fluid-gravity dualities and explore
the issue of entropy currents in those contexts. Another possible direction of
further research would be to include background fields along the lines of
\cite{Hur:2008tq,Kharzeev:2011ds,Loganayagam:2011mu}.

\begin{acknowledgments}
The authors would like to thank Micha\l\ P. Heller for discussions and helpful
comments on the manuscript, and Kasper
Peeters for his excellent package {\tt Cadabra}
\cite{DBLP:journals/corr/abs-cs-0608005,Peeters:2007wn}. This work was
partially supported by Polish Ministry of Science and Higher Education grant
\emph{N N202 173539}.
\end{acknowledgments}

\appendix

\section{Weyl covariance}
\label{appendixWeyl}

Conformal symmetry of ${\cal N}=4$ supersymmetric Yang-Mills theory 
can be extended to the bulk as follows
\cite{Bhattacharyya:2008ji,Bhattacharyya:2008mz}: 
\bel{Weylrescalings}
g_{\mu \nu} \rightarrow e^{- 2\phi} g_{\mu \nu}, \quad u^{\mu} \rightarrow
e^{\phi} u^{\mu}, \ \quad b \rightarrow e^{-\phi} b \quad \mathrm{and} \quad r
\rightarrow e^{\phi} r \, ,
\ee
where $\phi$ depends on the coordinates $x^{\mu}$
\cite{Bhattacharyya:2008mz}. A quantity which transforms homogeneously with a
factor of $e^{w\phi}$ is said to transform with Weyl weight $w$. 

A beautiful formalism allowing for manifest Weyl covariance in conformal
hydrodynamics was introduced by Loganayagam \cite{Loganayagam:2008is} and
applied to fluid-gravity duality in
\cite{Bhattacharyya:2008ji,Bhattacharyya:2008mz}. The basic tool is the 
Weyl-covariant derivative $\DD_{\mu}$, which preserves the Weyl weight
of the differentiated tensor. It is constructed using the vector field 
$\WA_{\nu}$ defined by \cite{Loganayagam:2008is}
\bel{connectionFIELD}
\WA_{\nu} \equiv u^\lambda\nabla_\lambda u_{\nu}-
\frac{\nabla_\lambda  u^\lambda}{3} u_{\nu} \, .
\ee
This quantity is of order one in the gradient expansion and transforms as a
connection under Weyl-transformations 
\bel{coonectionFIELDtrafo}
\WA_{\nu} \rightarrow \WA_{\nu} +\partial_{\nu}\phi \, .
\ee
Due to this property it can be used to compensate for derivatives of
the Weyl factor when differentiating a Weyl-covariant tensor. For instance,
one has
\bea
\DD_{\mu} b &=&  \partial_{\mu} b - \WA_{\mu} b \label{Db}, \\
\DD_{\mu} q &=&  \partial_{\mu} q + 3 \WA_{\mu} q \label{Dq}\, .
\eea
For further details the
reader is referred to the original literature cited above.

\section{Lorentz tensors at second order}
\label{appendixTensors}

\begin{itemize}
\item Scalars:
 \beal{scalars}
 \nonumber 
 S_1 &=& b^2 \sigma_{\mu \nu} \sigma^{\mu \nu},
 \\[0ex]
 \nonumber
 S_2 &=& b^2 \omega_{\mu \nu} \omega^{\mu \nu},
 \\[0ex]
 \nonumber
 S_3  &=& b^2 {\cal{R}},
 \\[0ex]
 \nonumber
 S_4 &=& b^2 q^{-2} P^{\mu \nu} \DD_{\mu}{q} \DD_{\nu}{q},
 \\[0ex]
 \nonumber
 S_5 &=& b^2 q^{-1} P^{\mu \nu} \DD_{\mu} \DD_{\nu}{q},
 \\[0ex]
 S_6 &=& b^2 q^{-1} P^{\mu \nu} l_{\mu} \DD_{\nu}{q} \, .
 \eea
 \item  Vectors:
 \bea
 \nonumber
 {V_1}_{\mu} &=& b P_{\mu \nu} \DD_{\rho} \sigma^{\nu \rho},
 \\[0ex]
 \nonumber
 {V_2}_{\mu} &=& b P_{\mu \nu} \DD_{\rho} \omega^{\nu \rho},
 \\[0ex]
 \nonumber
 {V_3}_{\mu} &=& b l^{\lambda} \sigma_{\mu \lambda},
 \\[0ex]
 \nonumber
 {V_4}_{\mu} &=& b q^{-1} \sigma_{\mu}^{\, \, \, \alpha} \DD_{\alpha}{q},
 \\[0ex]
  {V_5}_{\mu} &=& b q^{-1} \omega_{\mu}^{\, \, \, \alpha} \DD_{\alpha}{q} \, .
 \eea
 \item Tensors:
 \bea
 \nonumber
 {T_1}_{\mu \nu} &=& u^{\rho} \DD_{\rho} \sigma_{\mu \nu},
 \\[0ex]
 \nonumber
 {T_2}_{\mu \nu} &=& C_{\mu \alpha \nu \beta} u^{\alpha} u^{\beta},
 \\[0ex]
 \nonumber
 {T_3}_{\mu \nu} &=& \omega_{\mu}^{\, \, \, \lambda} \sigma_{\lambda \nu}+\omega_{\nu}^{\, \, \, \lambda} \sigma_{\lambda \mu},
 \\[0ex]
 \nonumber
 {T_4}_{\mu \nu} &=& \sigma_{\mu}^{\, \, \, \lambda} \sigma_{\lambda \nu} - \frac{1}{3} P_{\mu \nu} \sigma_{\alpha \beta} \sigma^{\mu \nu},
 \\[0ex]
 \nonumber
 {T_5}_{\mu \nu} &=& \omega_{\mu}^{\, \, \, \lambda} \omega_{\lambda \nu} + \frac{1}{3} P_{\mu \nu} \omega_{\alpha \beta} \omega^{\alpha \beta},
 \\[0ex]
 \nonumber
  {T_6}_{\mu \nu} &=&  \Pi_{\mu \nu}^{\alpha \beta} \DD_{\alpha} l_{\beta},
  \\[0ex]
  \nonumber
 {T_7}_{\mu \nu} &=& \frac{1}{2} \epsilon^{\alpha \beta}_{ \,\,\,\,\,\, \lambda (\mu } C_{\alpha \beta \nu) \sigma} u^{\lambda} u^{\sigma},
 \\[0ex]
 \nonumber
 {T_8}_{\mu \nu} &=& q^{-2} \Pi_{\mu \nu}^{\alpha \beta} \DD_{\alpha}{q} \DD_{\beta}{q},
 \\[0ex]
 \nonumber
 {T_9}_{\mu \nu} &=& q^{-1} \Pi_{\mu \nu}^{\alpha \beta} \DD_{\alpha} \DD_{\beta}q,
 \\[0ex]
 \nonumber
 {T_{10}}_{\mu \nu} &=& q^{-1} \Pi_{\mu \nu}^{\alpha \beta} l_{\alpha} \DD_{\beta}{q},
 \\[0ex]
   {T_{11}}_{\mu \nu} &=& \frac{1}{2} \epsilon_{(\mu}^{\, \, \, \, \, \alpha \beta \lambda}   \sigma_{\nu) \lambda}
  u_{\alpha} q^{-1} \DD_{\beta}{q} \, .
 \eea
\end{itemize}
Here $ \Pi_{\mu \nu}^{\alpha \beta} $ is the projector which can be used to
create symmetric, traceless tensors: 
\be
\Pi_{\mu \nu}^{\alpha \beta} = \frac{1}{2} \left( P_{\mu}^{\alpha}
 P_{\nu}^{\beta}+P_{\nu}^{\alpha} P_{\mu}^{\beta}-\frac{2}{3} P^{\alpha \beta}
 P_{\mu \nu} \right)\, .
\ee
The  scalar $\cal{R}$ is defined as in \cite{Bhattacharyya:2008mz} and
$C_{\mu \alpha \nu \beta}$ denotes the Weyl tensor:  
\be
C_{\mu \nu \lambda \sigma} = R_{\mu \nu \lambda \sigma} - \frac{1}{d-2} \left(
h_{\mu [\lambda} R_{\sigma] \nu}-h_{\nu [\lambda}R_{\sigma] \mu}  \right)+ 
\frac{1}{(d-1)(d-2)}  h_{\mu [\lambda} h_{\sigma] \nu} R \, .
\ee

\section{Second order hydrodynamics}
\label{appendixHydro}

As shown in \cite{Plewa:2012vt} the validity of the gradient expanded solution requires
the conservation of 
\beal{emtensor}
 \nonumber
 T_{\mu \nu} &=& \frac{1}{16 \pi \GN} \Big( \, \frac{1+b^6 q^2}{b^4}(P_{\mu
   \nu}+3 u_{\mu} u_{\nu}) - \frac{2  \sigma_{\mu \nu}}{b^3}+ \frac{ 2 (1+c_1)
   {T_1}_{\mu \nu} }{b^2}+\frac{2 {T_2}_{\mu \nu}}{b^2}+\frac{2 c_1 \,
   {T_3}_{\mu \nu}}{b^2}+\nonumber\\
&+& \frac{2 {T_4}_{\mu \nu}}{b^2} 
+ \frac{4 b^4 q^2 (-1+b^6 q^2 (12
   \kappa^2-1))}{1+b^6 q^2} {T_5}_{\mu \nu}+ \frac{ 2 \sqrt{3} b^7 q^3 \kappa
 }{1+b^6 q^2} {T_6}_{\mu \nu} +\frac{c_8}{b^2} {T_8}_{\mu \nu} +\nonumber\\
&+&
 \frac{c_9}{b^2} {T_9}_{\mu \nu} + \frac{ c_{10}}{b^2} {T_{10}}_{\mu \nu} 
\, \Big) \, ,
 \eea
and
\beal{chargecurrent}
  J_\mu &=& \frac{1}{8 \pi \GN} \Big( \frac{\sqrt{3}q u_\mu}{2}
  + \frac{3 b^4 q^2 \kappa l_\mu}{2(1+b^6 q^2)}
  - \frac{\sqrt{3} b^3 q (2+ b^6 q^2)}{8 (1+b^6 q^2) b^2} {V_0}_\mu + \frac{3 \sqrt{3} b q }{8(1+b^6 q^2)}{V_1}_{\mu} +
 \nonumber\\[1ex]  
  &+&\frac{3 \sqrt{3}b^7 q^3 \kappa^2 }{(1+b^6 q^2)^2}{V_2}_{\mu} - \frac{3 b^4 \kappa  q^2}{2 \left(b^6 q^2+1\right)^2} {V_3}_{\mu}
  + \frac{2 {a_4} \left(b^6 q^2+1\right)+\sqrt{3} b^9 q^3}{16 b^2 \left(b^6
    q^2+1\right)^2} {V_4}_{\mu} +
 \nonumber\\[1ex]  
&+&
\frac{ {a_5} \left(b^6 q^2+1\right)+\sqrt{3} b^9 \left(24 \kappa ^2-1\right)
  q^3-\sqrt{3} b^3 q}{8 b^2 \left(b^6 q^2+1\right)^2} {V_5}_{\mu} 
  \Big) \, ,
 \eea
where all the quantities appearing above are given in \cite{Plewa:2012vt}.

This leads to the equations of hydrodynamics
\beal{hydro}
\nonumber
{\DD_\mu}{b} &=& \frac{b^7 q^2 {V_0}_{\mu }}{2-b^6 q^2} +  \frac{b {V_1}_{\mu }}{b^6 q^2-2}+ \frac{3 b^7 q^2 {V_4}_{\mu }}{\left(b^6 q^2-2\right)^2}+\frac{{S_1} u_{\mu }}{6-3 b^6 q^2}
-\frac{b^{12} q^4 {S_4} u_{\mu }}{4 \left(b^6 q^2-2\right) \left(b^6 q^2+1\right)}+ \\ \nonumber
&+&\frac{b^6 q^2 \left(b^6 q^2+2\right) {S_5}  u_{\mu }}{4 \left(b^6 q^2-2\right) \left(b^6 q^2+1\right)} 
-\frac{2 \sqrt{3} b^9 \kappa  q^3 {S_6} u_{\mu }}{\left(b^6 q^2-2\right) \left(b^6 q^2+1\right)},
\\
\DD_\mu q &=& q {V_0}_{\mu }+\frac{b^5 q^3 {S_4}  u_{\mu }}{4
   b^6 q^2+4} -\frac{q    \left(b^6 q^2+2\right) {S_5} u_{\mu }}{4 \left(b^7 q^2+b\right)}+\frac{2 \sqrt{3} b^2 \kappa  q^2 {S_6}   u_{\mu }}{b^6 q^2+1}\, .
\eea
From these it follows that
\bea
\nonumber
\partial_\mu \WA_\nu - \partial_\nu \WA_\mu &=& \frac{b^6 q^2}{2 - b^6 q^2 } (
\DD_\nu {V_0}_\mu - \DD_\mu {V_0}_\nu  )\, ,
\\
\partial_\mu \WA_\nu - \partial_\nu \WA_\mu &=& -\frac{1}{3} (  \DD_\nu {V_0}_\mu - \DD_\mu {V_0}_\nu  ) \, ,
\eea
which leads to the relations \rf{fromhydro} used earlier.

\section{The expansions}
\label{appendixThetas}

The coefficients appearing in the expansion $\tl$ are\footnote{The prime
  denotes a derivative with respect to $br$.}:
\bea
\nonumber
\theta_l^1 &=& 3 b r {K_1}-2 b^2 B r^2 {F_2} {F_2}'+\frac{3}{2} b^2 B r^2 {L_1}',
\\[1ex]
\nonumber
\theta_l^2 &=& \frac{5 b^6 q^2 r^2+b^4 \left(4 r^6-7 q^2\right)+5 r^2}{4 b^5 r^7} + 
\frac{3 b^3 \kappa ^2 q^4}{5 r^9 \left(b^6 q^2+1\right)^2 \left(b^4 q^2-b^2 r^4-r^2\right)} \Big(
\\[1ex]
\nonumber
&+& 10 b^{20} q^4 r^{10}-5 b^{18} q^4 r^8-5 b^{16} q^4 r^6+5 b^{14} q^2 r^4 \left(q^2+2 r^6\right)+5 b^{12} q^2 r^2 \left(q^2-4 r^6\right)+
\\[1ex]
\nonumber
&-&4 b^{10} \left(2 q^4+5 q^2 r^6\right)+12 b^8 q^2 r^4+9 b^6 \left(2 q^2 r^2-r^8\right)-b^4 \left(8
   q^2+9 r^6\right)+13 b^2 r^4+13 r^2 
\Big),
\\[1ex]
\nonumber
\theta_l^3 &=&\frac{1}{4 b r},
\\[1ex]
\nonumber
\theta_l^4 &=&\frac{b^7 q^2}{32 r \left(b^6 q^2-2\right)^3 \left(b^6 q^2+1\right)^2 (1-b^2 r^2)(-b^6 q^2+b^2 r^2+b^4 r^4)} \Big( 
\\[1ex]
\nonumber
&&18 b^8 q^2 r^4 \left(b^6 q^2-2\right) \left(b^6 q^2+1\right) \left(b^6 q^2+2\right)^2
-12 b r^3 \left(b^6 q^2+1\right)^2 \left(b^{18} q^6+14 b^{12} q^4+12 b^6 q^2+8\right)+
\\[1ex]
\nonumber
&-&32 b^4 r^6 \left(b^6 q^2+1\right)^2 \left(b^{12} q^4-7 b^6
   q^2-2\right) 
   + 16 b^4 \left(b^6 q^3+q\right)^2 \left(b^{12} q^4+8 b^6 q^2+4\right)+
\\[1ex]
\nonumber   
   &-&r^2 \left(b^6 q^2 \left(9 b^{24} q^8+214 b^{18} q^6+580 b^{12} q^4+456 b^6 q^2+64\right)+64\right)+
\\[1ex]
\nonumber   
   &+& 12 b^5 r^7 \left(b^6 q^2+1\right) \left(b^{18}
   q^6+14 b^{12} q^4+12 b^6 q^2+8\right)+
\\[1ex]
\nonumber   
   &-&12 b^5 q^2 r \left(b^6 q^2+1\right) \left(b^{18} q^6-18 b^{12} q^4-20 b^6 q^2+8\right)
\Big)+
\\[1ex]
\nonumber
&+& \frac{b^2 {F_0}}{4 \left(b^6 q^2-2\right) \left(b^6 q^2+1\right)(1-b^2 r^2) (-b^6 q^2+b^2 r^2+b^4 r^4 )}
\Big(2 b^{18} q^6 r^2-3 b^{17} q^6 r+
\\[1ex]
\nonumber
&+& 2 b^{16} \left(q^6-5 q^4 r^6\right)+ 6 b^{10} \left(q^4-q^2 r^6\right)-6 b^6 q^2 r^2+12 b^5 q^2 r+4 b^4 \left(q^2+r^6\right)-4 r^2
\Big)+
\\[1ex]
\nonumber
&+& 3 b r {K_4}+\frac{b^3 q \left(b^6 q^2+1\right) }{b^6 q^2-2} \frac{\partial F_0}{\partial ( b^3 q)} +\frac{3}{2} b^2 B r^2 {L_4}'+\frac{b^3 r^3 \left(b^6 q^2-3 b^4 r^4+1\right)
   {F_0}^2}{2 (1-b^2 r^2) \left(-b^6 q^2 + b^2 r^2+b^4 r^4\right)},
   \\[1ex]
   \nonumber
  \theta_l^5 &=&-\frac{1}{2} {F_0}+3 b r {K_5}+\frac{3}{2} b^2 B r^2 {L_5}'-\frac{b^5 q^2 \left(3 b^7 q^2 r+4 b^6 q^2+6 b r+4\right)}{8 r \left(b^6 q^2-2\right)
   \left(b^6 q^2+1\right)},
   \\[1ex]
   \nonumber
  \theta_l^6 &=&\frac{3 \sqrt{3} b^3 \kappa  q^3}{4 r^6 \left(b^6 q^2-2\right) \left(b^6 q^2+1\right)^2 \left(b^4 q^2-b^2 r^4-r^2\right)} \Big(
 2 b^{23} q^6 r^7+b^{17} q^4 r \left(q^2+6 r^6\right)+4 b^{16} q^4 r^6-3 b^{15} q^4 r^5
\\[1ex]
\nonumber 
 &-&4 b^{14} q^2 r^{10}-3 b^{13} q^4 r^3-4 b^{12} q^2 r^8+4 b^{11} q^2 r^7+4 b^{10} q^2 \left(q^2+r^6\right)-6 b^9 q^2 r^5-4 b^8 r^4
   \left(q^2+r^6\right)+
\\[1ex]   
   &-&6 b^7 q^2 r^3-4 b^6 r^2 \left(q^2+r^6\right)-4 b^5 q^2 r+4 b^4 q^2-4 b^2 r^4-4 r^2
  \Big)  +
  3 b r {K_6}+\frac{3}{2} b^2 B r^2 {L_6}' \, .
\eea
For $\theta_n$ the result is
\bea
\nonumber
\theta^1_n &=&2 {F_2} {F_2}'-\frac{3}{2} {L_1}',
\\[1ex]
\nonumber
\theta^2_n &=&\frac{1}{b^3 r^3} + \frac{6 b^2 \kappa ^2 q^4}{5 r^7 \left(b^6 q^2+1\right)^2 \left(-b^4 q^2 +b^2 r^4+r^2\right)^2} \Big(
\\[1ex]
\nonumber
&& b^3 (9 b^8 q^4+10 b^{12} q^2 r^{10} (b^6 q^2+1)+r^4 (5 b^{12} q^4-28 b^6 q^2-6)+
\\[1ex]
\nonumber
&+&b^4 r^8 (5 b^6 q^2 (b^6 q^2-2)-6)+2 b^4 q^2 r^2 (5 b^6 q^2-4)-6 b^2 r^6 (5 b^6
   q^2+2))
\Big),
\\[1ex]
\nonumber
\theta^3_n &=&0,
\\[1ex]
\nonumber
\theta^4_n &=&\frac{b^7 q^2 r^3}{16 \left(b^6 q^2-2\right)^3 \left(b^6 q^2+1\right)^2 \left(b^2 r^2-1\right)^2 \left(-b^4 q^2+b^2 r^4+r^2\right)^2} \Big(
\\[1ex]
\nonumber
&& 18 b^8 q^2 r^4 \left(b^6 q^2-2\right) \left(b^6 q^2+1\right) \left(b^6 q^2+2\right)^2
+12 b r^3 \left(b^6 q^2+1\right)^2 \left(b^{18} q^6+14 b^{12} q^4+12 b^6 q^2+8\right)+
\\[1ex]
\nonumber
&+& 16 b^4 r^6 \left(b^6 q^2+1\right)^2 \left(b^{12} q^4+8 b^6
   q^2+4\right)
   +32 b^4 q^2 \left(b^6 q^2+1\right)^2 \left(2 b^{12} q^4+b^6 q^2+2\right)+
\\[1ex]
\nonumber   
   &-& r^2 \left(b^6 q^2 \left(57 b^{24} q^8+262 b^{18} q^6+436 b^{12} q^4+216 b^6 q^2-32\right)+64\right)+
\\[1ex]
\nonumber   
   &-& 12 b^5 r^7 \left(b^6 q^2+1\right) \left(b^{18}
   q^6+14 b^{12} q^4+12 b^6 q^2+8\right)+
\\[1ex]
\nonumber   
   &-& 12 b^5 q^2 r \left(b^6 q^2+1\right) \left(3 b^{18} q^6+10 b^{12} q^4+4 b^6 q^2+24\right)
\Big)+
\\[1ex]
\nonumber
&-& \frac{b^2 r^4 {F_0}}{2 \left(b^6 q^2-2\right) \left(b^6 q^2+1\right) \left(b^6 q^2 r^2-b^4 \left(q^2+r^6\right)+r^2\right)^2}
\Big(
-2 b^4 q^2 (2 + 9 b^6 q^2 + 7 b^{12} q^4) +
\\[1ex]
\nonumber 
  &+&
 3 b^5 q^2 (-4 + b^{12} q^4) r  -2 b^4 (2 + 3 b^6 q^2 + b^{12} q^4) r^6 + 
 2 (2 + 5 b^6 q^2) (r + b^6 q^2 r)^2
\Big) +
\\[1ex]
\nonumber
&-& \frac{b^3 r^5 \left(2 b^6 q^2 r^2-3 b^4 q^2+2 r^2\right) {F_0}^2}{\left(b^6 q^2 r^2-b^4 \left(q^2+r^6\right)+r^2\right)^2}+
\frac{2 b^5 q r^4 \left(b^6 q^2+1\right) }{\left(b^6 q^2-2\right) \left(b^2 r^2-1\right) \left(b^4 q^2-b^2 r^4-r^2\right)} \frac{\partial F_0}{\partial (q b^3)}  -\frac{3}{2} {L_4}',
\\[1ex]
\nonumber
\theta^5_n &=&-\frac{3}{2} {L_5}'+\frac{{F_0}}{2 b^2 B r^2}+\frac{b^3 q^2 \left(-3 b^7 q^2 r+4 b^6 q^2-6 b r+4\right)}{8 B r^3 \left(b^6 q^2-2\right) \left(b^6 q^2+1\right)},
\\[1ex]
\nonumber
\theta^6_n &=&\frac{\sqrt{3} b^5 \kappa  q^3}{2 r^2 \left(b^6 q^2-2\right) \left(b^6 q^2+1\right) \left(-b^4 q^2 +b^2 r^4+r^2\right)^2} \Big(
6 b^{15} q^4 r^5+3 b^{13} q^4 r^3+4 b^{12} q^4 r^2+8 b^{10} q^4+
\\[1ex]
\nonumber
&+& 12 b^9 q^2 r^5-12 b^8 q^2 r^4+6 b^7 q^2 r^3+b^6 \left(12 r^8-8 q^2 r^2\right)-4 b^4 \left(q^2-6 r^6\right)+24 b^2 r^4+12 r^2
\Big)+
\\[1ex]
\nonumber
&+& \frac{\sqrt{3} b^6 \kappa  q^3}{2 r^2 \left(b^6 q^2+1\right)^2 \left(b^2 r^2-1\right) \left(-b^4 q^2+b^2 r^4+r^2\right)^2} \Big(
\\[1ex]
\nonumber
&& 12 r \left(-b^{10} q^4+b^8 q^2 r^4+b^6 q^2 r^2-b^4 q^2+b^2 r^4+r^2\right) {F_0}+
\\[1ex]
&+& b^3 q^2 \left(b^4 r^4+b^2 r^2+1\right) \left(3 b^7 q^2 r-4 b^6 q^2+6 b r-4\right)
\Big)  -\frac{3}{2} {L_6}' \, .
\eea

\section{The entropy current at first order}
\label{appendixFO}

The authors of reference \cite{Son:2009tf} considered the most general form of
entropy current and constitutive relations in hydrodynamics with a charge
current (allowing for the possibility of a $U(1)$ anomaly). They showed that
at first order in gradients 
the entropy current must be of the form
\bel{echydro}
S^\mu = s u^\mu  - \frac{\mu}{T} \nu^\mu + \half D l^\mu \, ,
\ee
where $s$ is the equilibrium entropy density, $D$ is a coefficient
discussed below, and $\nu^\mu$ is the first order 
correction to the charge current:
\be
J^\mu = n u^\mu + \nu^\mu \, .
\ee
By symmetry arguments this correction can be expressed as
\be
\nu^\mu = - \mu \sigma P^{\mu \nu} \partial_\nu \left( \frac{\mu}{T}\right) +  \half
\xi l^\mu \, ,
\ee
where $\sigma$ and
$\xi$ are transport coefficients. 
For the fluid considered in this paper these transport coefficients 
can be read off from \rf{chargecurrent}. To do this one needs the relations 
\bel{Tmu}
T =  \frac{2-b^6 q^2}{2 \pi  b}, \quad \mu = \frac{\sqrt{3} b^2 q}{\pi }\, ,
\ee
from which one derives
\be
\label{toV0}
P^{\mu \nu} \partial_\nu \left( \frac{\mu}{T} \right) = -\frac{4 \sqrt{3} b^3
  q \left(b^{12} q^4+3 b^6 q^2+2\right)}{\left(b^6
  q^2-2\right)^3}  {V_0}^{\mu }\, .
\ee
The coefficient $\xi$ is non-vanishing only if the theory is anomalous (in the
sense that coupling it to a background gauge fields results in an anomalous
divergence of the charge current). As shown by Son and Sur\'owka
\cite{Son:2009tf}, $\xi$ is proportional to the anomaly coefficient $C$:
\be
\xi = C \left(\mu^2 - \frac{2}{3} \frac{n \mu^3}{\epsilon + p}\right) \, .
\ee
This equation allows one to express $C$ in terms of known quantities, and then 
one can compute $D$ using the relation \cite{Son:2009tf}
\be
\label{DSon}
D = \frac{1}{3} C \frac{\mu^3}{T} \, .
\ee
Proceeding in this fashion one can express the entropy current \rf{echydro} in
terms of the variables $b$ and 
$q$ used in this paper. This leads to
\be
S^\mu  =  \frac{1}{4 \GN} \left( \frac{1}{b^3} u^\mu - \frac{3 b^4 q^2 \left(b^6
  q^2+2\right)}{4 \left(b^6 q^2-2\right) \left(b^6 q^2+1\right)} V_0^\mu  
-\frac{\sqrt{3} b^7 \kappa  q^3}{\left(b^6  q^2+1\right)}  l^\mu\right) \, .
\ee
This is precisely the first order part of \rf{ecres} with coefficients
  \rf{econe}, which were derived from the holographic entropy current formula
  \rf{entrocur} proposed in \cite{Booth:2010kr}.

\bibliographystyle{utphys}
\bibliography{biblio.bib}

\providecommand{\href}[2]{#2}\begingroup\raggedright\begin{thebibliography}{10}

\bibitem{Booth:2010kr}
I.~Booth, M.~P. Heller, and M.~Spalinski, ``{Black Brane Entropy and
  Hydrodynamics},'' \href{http://dx.doi.org/10.1103/PhysRevD.83.061901}{{\em
  Phys. Rev.} {\bfseries D83} (2011) 061901},
\href{http://arxiv.org/abs/1010.6301}{{\ttfamily arXiv:1010.6301 [hep-th]}}.

\bibitem{Booth:2011qy}
I.~Booth, M.~P. Heller, G.~Plewa, and M.~Spalinski, ``{On the apparent horizon
  in fluid-gravity duality},''
  \href{http://dx.doi.org/10.1103/PhysRevD.83.106005}{{\em Phys.Rev.}
  {\bfseries D83} (2011) 106005},
\href{http://arxiv.org/abs/1102.2885}{{\ttfamily arXiv:1102.2885 [hep-th]}}.

\bibitem{Son:2009tf}
D.~T. Son and P.~Surowka, ``{Hydrodynamics with Triangle Anomalies},''
  \href{http://dx.doi.org/10.1103/PhysRevLett.103.191601}{{\em Phys. Rev.
  Lett.} {\bfseries 103} (2009) 191601},
\href{http://arxiv.org/abs/0906.5044}{{\ttfamily arXiv:0906.5044 [hep-th]}}.

\bibitem{Bekenstein:1973ur}
J.~D. Bekenstein, ``{Black holes and entropy},''
\href{http://dx.doi.org/10.1103/PhysRevD.7.2333}{{\em Phys.Rev.} {\bfseries D7}
  (1973) 2333--2346}.

\bibitem{Booth:2003ji}
I.~Booth and S.~Fairhurst, ``{The first law for slowly evolving horizons},''
  \href{http://dx.doi.org/10.1103/PhysRevLett.92.011102}{{\em Phys. Rev. Lett.}
  {\bfseries 92} (2004) 011102},
\href{http://arxiv.org/abs/gr-qc/0307087}{{\ttfamily arXiv:gr-qc/0307087}}.

\bibitem{Booth:2005qc}
I.~Booth, ``{Black hole boundaries},''
  \href{http://dx.doi.org/10.1139/p05-063}{{\em Can. J. Phys.} {\bfseries 83}
  (2005) 1073--1099},
\href{http://arxiv.org/abs/gr-qc/0508107}{{\ttfamily arXiv:gr-qc/0508107}}.

\bibitem{Figueras:2009iu}
P.~Figueras, V.~E. Hubeny, M.~Rangamani, and S.~F. Ross, ``{Dynamical black
  holes and expanding plasmas},''
  \href{http://dx.doi.org/10.1088/1126-6708/2009/04/137}{{\em JHEP} {\bfseries
  04} (2009) 137},
\href{http://arxiv.org/abs/0902.4696}{{\ttfamily arXiv:0902.4696 [hep-th]}}.

\bibitem{Booth:2009ct}
I.~Booth, M.~P. Heller, and M.~Spalinski, ``{Black brane entropy and
  hydrodynamics: the boost-invariant case},''
  \href{http://dx.doi.org/10.1103/PhysRevD.80.126013}{{\em Phys. Rev.}
  {\bfseries D80} (2009) 126013},
\href{http://arxiv.org/abs/0910.0748}{{\ttfamily arXiv:0910.0748 [hep-th]}}.

\bibitem{Bhattacharyya:2008jc}
S.~Bhattacharyya, V.~E. Hubeny, S.~Minwalla, and M.~Rangamani, ``{Nonlinear
  Fluid Dynamics from Gravity},''
  \href{http://dx.doi.org/10.1088/1126-6708/2008/02/045}{{\em JHEP} {\bfseries
  02} (2008) 045},
\href{http://arxiv.org/abs/0712.2456}{{\ttfamily arXiv:0712.2456 [hep-th]}}.

\bibitem{Bhattacharyya:2008xc}
S.~Bhattacharyya {\em et al.}, ``{Local Fluid Dynamical Entropy from
  Gravity},'' {\em JHEP} {\bfseries 06} (2008) 055,
\href{http://arxiv.org/abs/0803.2526}{{\ttfamily arXiv:0803.2526 [hep-th]}}.

\bibitem{Loganayagam:2008is}
R.~Loganayagam, ``{Entropy Current in Conformal Hydrodynamics},''
  \href{http://dx.doi.org/10.1088/1126-6708/2008/05/087}{{\em JHEP} {\bfseries
  05} (2008) 087},
\href{http://arxiv.org/abs/0801.3701}{{\ttfamily arXiv:0801.3701 [hep-th]}}.

\bibitem{LL}
L.~D. Landau and E.~M. Lifshitz, {\em Fluid Mechanics, Second Edition: Volume 6
  (Course of Theoretical Physics)}.
\newblock Butterworth-Heinemann, 2~ed., January, 1987.

\bibitem{Romatschke:2009im}
P.~Romatschke, ``{New Developments in Relativistic Viscous Hydrodynamics},''
  \href{http://dx.doi.org/10.1142/S0218301310014613}{{\em Int. J. Mod. Phys.}
  {\bfseries E19} (2010) 1--53},
\href{http://arxiv.org/abs/0902.3663}{{\ttfamily arXiv:0902.3663 [hep-ph]}}.

\bibitem{Bhattacharyya:2012nq}
S.~Bhattacharyya, ``{Constraints on the second order transport coefficients of
  an uncharged fluid},'' \href{http://dx.doi.org/10.1007/JHEP07(2012)104}{{\em
  JHEP} {\bfseries 1207} (2012) 104},
\href{http://arxiv.org/abs/1201.4654}{{\ttfamily arXiv:1201.4654 [hep-th]}}.

\bibitem{Compere:2012mt}
G.~Compere, P.~McFadden, K.~Skenderis, and M.~Taylor, ``{The relativistic fluid
  dual to vacuum Einstein gravity},''
  \href{http://dx.doi.org/10.1007/JHEP03(2012)076}{{\em JHEP} {\bfseries 1203}
  (2012) 076},
\href{http://arxiv.org/abs/1201.2678}{{\ttfamily arXiv:1201.2678 [hep-th]}}.

\bibitem{Erdmenger:2008rm}
J.~Erdmenger, M.~Haack, M.~Kaminski, and A.~Yarom, ``{Fluid dynamics of
  R-charged black holes},''
  \href{http://dx.doi.org/10.1088/1126-6708/2009/01/055}{{\em JHEP} {\bfseries
  0901} (2009) 055}, \href{http://arxiv.org/abs/0809.2488}{{\ttfamily
  arXiv:0809.2488 [hep-th]}}.

\bibitem{Banerjee:2008th}
N.~Banerjee, J.~Bhattacharya, S.~Bhattacharyya, S.~Dutta, R.~Loganayagam, {\em
  et al.}, ``{Hydrodynamics from charged black branes},''
  \href{http://dx.doi.org/10.1007/JHEP01(2011)094}{{\em JHEP} {\bfseries 1101}
  (2011) 094}, \href{http://arxiv.org/abs/0809.2596}{{\ttfamily arXiv:0809.2596
  [hep-th]}}.

\bibitem{Hur:2008tq}
J.~Hur, K.~K. Kim, and S.-J. Sin, ``{Hydrodynamics with conserved current from
  the gravity dual},''
  \href{http://dx.doi.org/10.1088/1126-6708/2009/03/036}{{\em JHEP} {\bfseries
  0903} (2009) 036},
\href{http://arxiv.org/abs/0809.4541}{{\ttfamily arXiv:0809.4541 [hep-th]}}.

\bibitem{Kalaydzhyan:2010iv}
T.~Kalaydzhyan and I.~Kirsch, ``{Holographic dual of a boost-invariant plasma
  with chemical potential},''
  \href{http://dx.doi.org/10.1007/JHEP02(2011)053}{{\em JHEP} {\bfseries 1102}
  (2011) 053},
\href{http://arxiv.org/abs/1012.1966}{{\ttfamily arXiv:1012.1966 [hep-th]}}.

\bibitem{Plewa:2012vt}
G.~Plewa and M.~Spalinski, ``{On the gravity dual of strongly coupled charged
  plasma},'' \href{http://dx.doi.org/10.1007/JHEP05(2013)002}{{\em JHEP}
  {\bfseries 1305} (2013) 002},
\href{http://arxiv.org/abs/1212.2344}{{\ttfamily arXiv:1212.2344 [hep-th]}}.

\bibitem{Chamblin:1999tk}
A.~Chamblin, R.~Emparan, C.~V. Johnson, and R.~C. Myers, ``{Charged AdS black
  holes and catastrophic holography},''
  \href{http://dx.doi.org/10.1103/PhysRevD.60.064018}{{\em Phys.Rev.}
  {\bfseries D60} (1999) 064018},
\href{http://arxiv.org/abs/hep-th/9902170}{{\ttfamily arXiv:hep-th/9902170
  [hep-th]}}.

\bibitem{Gauntlett:2006ai}
J.~P. Gauntlett, E.~O~Colgain, and O.~Varela, ``{Properties of some conformal
  field theories with M-theory duals},''
  \href{http://dx.doi.org/10.1088/1126-6708/2007/02/049}{{\em JHEP} {\bfseries
  0702} (2007) 049},
\href{http://arxiv.org/abs/hep-th/0611219}{{\ttfamily arXiv:hep-th/0611219
  [hep-th]}}.

\bibitem{Gauntlett:2007ma}
J.~P. Gauntlett and O.~Varela, ``{Consistent Kaluza-Klein reductions for
  general supersymmetric AdS solutions},''
  \href{http://dx.doi.org/10.1103/PhysRevD.76.126007}{{\em Phys.Rev.}
  {\bfseries D76} (2007) 126007},
\href{http://arxiv.org/abs/0707.2315}{{\ttfamily arXiv:0707.2315 [hep-th]}}.

\bibitem{Bhattacharyya:2008mz}
S.~Bhattacharyya, R.~Loganayagam, I.~Mandal, S.~Minwalla, and A.~Sharma,
  ``{Conformal Nonlinear Fluid Dynamics from Gravity in Arbitrary
  Dimensions},'' \href{http://dx.doi.org/10.1088/1126-6708/2008/12/116}{{\em
  JHEP} {\bfseries 12} (2008) 116},
\href{http://arxiv.org/abs/0809.4272}{{\ttfamily arXiv:0809.4272 [hep-th]}}.

\bibitem{deHaro:2000xn}
S.~de~Haro, S.~N. Solodukhin, and K.~Skenderis, ``{Holographic reconstruction
  of spacetime and renormalization in the AdS/CFT correspondence},''
  \href{http://dx.doi.org/10.1007/s002200100381}{{\em Commun. Math. Phys.}
  {\bfseries 217} (2001) 595--622},
\href{http://arxiv.org/abs/hep-th/0002230}{{\ttfamily arXiv:hep-th/0002230}}.

\bibitem{Booth:2006bn}
I.~Booth and S.~Fairhurst, ``{Isolated, slowly evolving, and dynamical trapping
  horizons: geometry and mechanics from surface deformations},''
  \href{http://dx.doi.org/10.1103/PhysRevD.75.084019}{{\em Phys. Rev.}
  {\bfseries D75} (2007) 084019},
\href{http://arxiv.org/abs/gr-qc/0610032}{{\ttfamily arXiv:gr-qc/0610032}}.

\bibitem{Romatschke:2009kr}
P.~Romatschke, ``{Relativistic Viscous Fluid Dynamics and Non-Equilibrium
  Entropy},'' \href{http://dx.doi.org/10.1088/0264-9381/27/2/025006}{{\em
  Class. Quant. Grav.} {\bfseries 27} (2010) 025006},
\href{http://arxiv.org/abs/0906.4787}{{\ttfamily arXiv:0906.4787 [hep-th]}}.

\bibitem{Sadofyev:2010pr}
A.~Sadofyev and M.~Isachenkov, ``{The Chiral magnetic effect in hydrodynamical
  approach},'' \href{http://dx.doi.org/10.1016/j.physletb.2011.02.041}{{\em
  Phys.Lett.} {\bfseries B697} (2011) 404--406},
\href{http://arxiv.org/abs/1010.1550}{{\ttfamily arXiv:1010.1550 [hep-th]}}.

\bibitem{Kanitscheider:2009as}
I.~Kanitscheider and K.~Skenderis, ``{Universal hydrodynamics of non-conformal
  branes},'' \href{http://dx.doi.org/10.1088/1126-6708/2009/04/062}{{\em JHEP}
  {\bfseries 0904} (2009) 062},
  \href{http://arxiv.org/abs/0901.1487}{{\ttfamily arXiv:0901.1487 [hep-th]}}.

\bibitem{Bhattacharya:2011eea}
J.~Bhattacharya, S.~Bhattacharyya, and S.~Minwalla, ``{Dissipative Superfluid
  dynamics from gravity},''
  \href{http://dx.doi.org/10.1007/JHEP04(2011)125}{{\em JHEP} {\bfseries 1104}
  (2011) 125},
\href{http://arxiv.org/abs/1101.3332}{{\ttfamily arXiv:1101.3332 [hep-th]}}.

\bibitem{Herzog:2011ec}
C.~P. Herzog, N.~Lisker, P.~Surowka, and A.~Yarom, ``{Transport in holographic
  superfluids},'' \href{http://dx.doi.org/10.1007/JHEP08(2011)052}{{\em JHEP}
  {\bfseries 1108} (2011) 052},
\href{http://arxiv.org/abs/1101.3330}{{\ttfamily arXiv:1101.3330 [hep-th]}}.

\bibitem{Kharzeev:2011ds}
D.~E. Kharzeev and H.-U. Yee, ``{Anomalies and time reversal invariance in
  relativistic hydrodynamics: the second order and higher dimensional
  formulations},'' \href{http://dx.doi.org/10.1103/PhysRevD.84.045025}{{\em
  Phys.Rev.} {\bfseries D84} (2011) 045025},
\href{http://arxiv.org/abs/1105.6360}{{\ttfamily arXiv:1105.6360 [hep-th]}}.

\bibitem{Loganayagam:2011mu}
R.~Loganayagam, ``{Anomaly Induced Transport in Arbitrary Dimensions},''
\href{http://arxiv.org/abs/1106.0277}{{\ttfamily arXiv:1106.0277 [hep-th]}}.

\bibitem{DBLP:journals/corr/abs-cs-0608005}
K.~Peeters, ``A field-theory motivated approach to symbolic computer algebra,''
  {\em CoRR} {\bfseries abs/cs/0608005} (2006) .

\bibitem{Peeters:2007wn}
K.~Peeters, ``{Introducing Cadabra: A symbolic computer algebra system for
  field theory problems},''
\href{http://arxiv.org/abs/hep-th/0701238}{{\ttfamily arXiv:hep-th/0701238}}.

\bibitem{Bhattacharyya:2008ji}
S.~Bhattacharyya {\em et al.}, ``{Forced Fluid Dynamics from Gravity},''
  \href{http://dx.doi.org/10.1088/1126-6708/2009/02/018}{{\em JHEP} {\bfseries
  02} (2009) 018},
\href{http://arxiv.org/abs/0806.0006}{{\ttfamily arXiv:0806.0006 [hep-th]}}.

\end{thebibliography}\endgroup

\end{document}